\documentclass[sigconf, screen]{acmart}
\renewcommand\footnotetextcopyrightpermission[1]{}  
\AtBeginDocument{%
  }

\setcopyright{acmlicensed}
\copyrightyear{2026}
\acmYear{2026}
\acmDOI{XXXXXXX.XXXXXXX}
\acmConference[MM '26]{34th ACM International Conference on Multimedia}{November 10--14,
  2026}{Rio de Janeiro, Brazil}
\acmConference[Arxiv]{Preprint}{Preprint}{2026}
\acmISBN{978-1-4503-XXXX-X/2018/06}

\usepackage{booktabs}
\usepackage{multirow}
\usepackage{booktabs}
\usepackage{multirow}
\usepackage{enumitem}
\usepackage{tabularx}
\usepackage{ragged2e}

\usepackage{amssymb} %
\usepackage{makecell}
\usepackage{tcolorbox}
\usepackage{listings}
\tcbuselibrary{breakable}
\begin{document}

\title{Listening with Time: Precise Temporal Awareness for Long-Form Audio Understanding}


\author{Mingchen Shao}
\affiliation{%
  \institution{Northwestern Polytechnical University}
  \city{Xi'an}
  \country{China}}
\email{mcshao@mail.nwpu.edu.cn}

\author{Hang Su}
\affiliation{%
  \institution{Independent Researcher}
  \city{Beijing}
  \country{China}}
\email{hangsucuhk@gmail.com}

\author{Wenjie Tian}
\affiliation{%
  \institution{Northwestern Polytechnical University}
  \city{Xi'an}
  \country{China}}
\email{twj@mail.nwpu.edu.cn}

\author{Bingshen Mu}
\affiliation{%
  \institution{Northwestern Polytechnical University}
  \city{Xi'an}
  \country{China}}
\email{bsmu@mail.nwpu.edu.cn}

\author{Zhennan Lin}
\affiliation{%
  \institution{Northwestern Polytechnical University}
  \city{Xi'an}
  \country{China}}
\email{znlin@mail.nwpu.edu.cn}

\author{Lichun Fan}
\affiliation{%
  \institution{Independent Researcher}
  \city{Beijing}
  \country{China}}
\email{fanlc2010@gmail.com}

\author{Zhenbo Luo}
\affiliation{%
  \institution{Independent Researcher}
  \city{Beijing}
  \country{China}}
\email{luozhenbo@tsinghua.org.cn}

\author{Jian Luan}
\affiliation{%
  \institution{Independent Researcher}
  \city{Beijing}
  \country{China}}
\email{jian.luan@gmail.com}

\author{Lei Xie}
\affiliation{%
  \institution{Northwestern Polytechnical University}
  \city{Xi'an}
  \country{China}}
\email{lxie@nwpu.edu.cn}







\renewcommand{\shortauthors}{Shao et al.}

\begin{abstract}
While Large Audio Language Models (LALMs) achieve strong performance on short audio, they degrade on long-form inputs.
This degradation is more severe in temporal awareness tasks, where temporal alignment becomes increasingly inaccurate as audio duration grows.
We attribute these limitations to the lack of data, benchmarks, and modeling approaches tailored for long-form temporal awareness. 
To bridge this gap, we first construct LAT-Chronicle, a 1.2k hour long-form audio dataset with temporal annotations across real-world scenarios.
We further develop LAT-Bench, the first human-verified benchmark supporting audio up to 30 minutes while covering three core tasks: Dense Audio Caption, Temporal Audio Grounding, and Targeted Audio Caption.
Leveraging these resources, we propose LAT-Audio, formulating temporal awareness as a progressive global-to-local reasoning paradigm.
A global timeline is first constructed as an aligned temporal-semantic context, and the Think-With-Audio Chain-of-Thought (TWA-CoT) is then introduced to perform iterative reasoning by incorporating local audio information via tool use.
Experiments show that LAT-Audio surpasses existing models on long-form audio temporal awareness tasks and improves robustness to input duration.
We release the dataset, benchmark, and model to facilitate future research at~\url{https://github.com/alanshaoTT/LAT-Audio-Repo}.

\end{abstract}

\maketitle
\section{Introduction}
Audio is a fundamental modality for real-world intelligent systems, conveying rich semantic, acoustic, and temporal information through diverse signals including speech, music, and sounds.
While Large Language Models (LLMs) have achieved remarkable success in language understanding and reasoning~\cite{phi,gpt4,deepseek,qwen3-llm-technical}, they are inherently limited to textual inputs.
Recent advances in Large Audio Language Models (LALMs) address this limitation by extending LLMs to the audio domain~\cite{gpt-4o, AF3,step-audio-r1,mimo-audio,qwen3omni}, enabling unified understanding and reasoning over diverse audio inputs.
\begin{figure}[t]
  \centering
  \includegraphics[width=\linewidth]{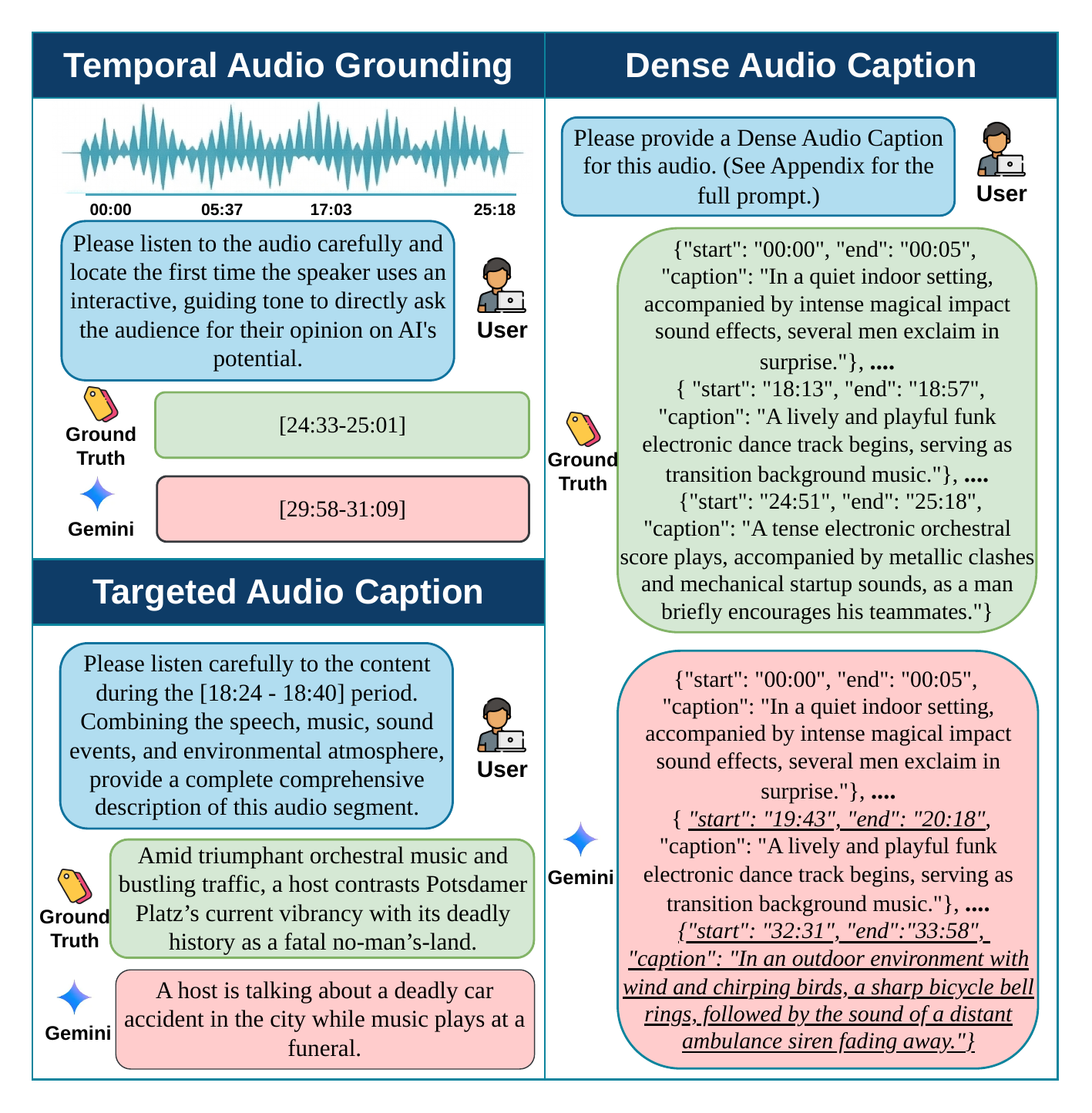}
  \caption{Examples of LATA tasks and typical failures: temporal hallucinations and timestamp drift.}
  \label{fig:数据示例和错误示例}
\end{figure}

However, most existing LALMs demonstrate strong audio understanding performance on short clips but exhibit significant degradation on long-form inputs, especially for Long-form Audio Temporal Awareness (LATA)~\cite{AF3, BLAB,longspeech}.
LATA tasks require models to jointly understand audio content and accurately localize events in time~\cite{trace,temporal-exp,huanjue}.
When handling LATA tasks, models often struggle to achieve accurate temporal alignment, exhibiting two typical failure patterns: temporal hallucination and timestamp drift.
Temporal hallucination refers to predicted events falling outside the valid temporal range, while timestamp drift denotes progressively deviating temporal alignment, both of which worsen with audio duration increase~\cite{BLAB,time-audio,tagspeech}, as illustrated in Fig.~\ref{fig:数据示例和错误示例}.
But in real-world scenarios, audio content, such as meetings, podcasts, recordings, and other media, typically spans several minutes to tens of minutes rather than short clips.
Although segmenting long audio into shorter chunks is a practical workaround, it inevitably disrupts the global context and breaks temporal continuity.

Research on LATA remains limited in datasets, benchmarks, and modeling approaches~\cite{lata-model,bench1,bench2,bench3}.
To the best of our knowledge, datasets specifically designed for LATA remain largely absent. 
Existing related datasets~\cite{AF3, time-audio,tacos} suffer from notable limitations, including a lack of precise, fine-grained temporal annotations, limited audio duration, and English-only content.
Moreover, existing benchmarks~\cite{BLAB, time-audio} either support long-form audio with limited task coverage, or provide temporal awareness tasks but are restricted to short clips.
This reveals a critical gap in comprehensive long-form audio benchmarking, which requires longer audio with diverse temporal grounding tasks and more diverse audio.
On the modeling side, only a few recent systems support long-form audio understanding~\cite{AF3, qwen3omni, gemini-2.5}.
However, they still struggle with LATA tasks, with performance degrading as the duration of input audio increases~\cite{trace,temporal-exp,qwen3-vl}.
This is mainly due to the large temporal context in long-form audio~\cite{whisperx,audiomarathon,speechXL,fastlm}, which makes accurate temporal localization difficult and leads to cumulative errors.
Unlike video, where temporal boundaries can often be inferred from observable visual transitions such as actions or scene changes~\cite{think-with-iamge,think-with-vedio,longvt}, audio events are typically continuous, overlapping, and weakly bounded, making precise temporal reasoning inherently more challenging.

To address these limitations, we propose a comprehensive solution across three aspects.
To address the scarcity of long-form audio with precise temporal annotations, we develop LAT-Pipe, a human-in-the-loop pipeline that generates multi-task temporal annotations across diverse audio.
Building upon this pipeline, we construct LAT-Chronicle, a 1.2k hour long-form dataset for LATA, covering diverse real-world audio scenarios in both Chinese and English.
To enable comprehensive and realistic evaluation, we introduce LAT-Bench, the first human-verified benchmark designed for long-form audio up to 30 minutes. 
It supports three core LATA tasks: Dense Audio Caption (DAC), Temporal Audio Grounding (TAG), and Targeted Audio Caption (TAC), along with corresponding evaluation metrics, and covers diverse and complex real-world audio scenarios.
Furthermore, we propose LAT-Audio, a framework that formulates LATA into a progressive global-to-local reasoning paradigm. 
The model first predicts a global timeline as an aligned temporal-semantic context, and then performs iterative reasoning via a Think-With-Audio Chain-of-Thought (TWA-CoT), where additional local audio information is introduced through tool use. 
By narrowing the temporal context and incorporating additional audio information, LAT-Audio achieves more precise temporal alignment and reduces temporal errors.
Experimental results show that it achieves state-of-the-art performance on TAG, DAC, and TAC tasks, improving robustness as input duration increases.

Our contributions can be summarized as follows:
\begin{itemize}
\item \textbf{Dataset:} We construct \textbf{LAT-Chronicle}, a 1.2kh long-form audio dataset with multi-dimensional temporal annotations across diverse real-world scenarios
in Chinese and English.
\item \textbf{Benchmark:} We develop \textbf{LAT-Bench}, the first
human-verified benchmark supporting audio up to 30 minutes while covering three core tasks: DAC, TAG, TAC.
\item \textbf{Framework:} We propose \textbf{LAT-Audio}, formulating
LATA as a progressive global-to-local reasoning paradigm with TWA-CoT, surpassing existing methods on LATA tasks and improving robustness to input duration.
\item \textbf{Open-source:} We open-source the dataset, benchmark, and model to fill this gap and facilitate future research on long-form audio temporal awareness.
\end{itemize}

\section{Related Work}
\subsection{Long-form Audio Temporal Awareness Resources}
The development of LATA in LALMs critically depends on datasets with precise temporal annotations and diverse audio sources. 
However, existing resources remain insufficient.
LongAudio-XL~\cite{AF3}, introduced with Audio-Flamingo 3, extends to audio durations of up to 10 minutes and covers multiple modalities, but lacks temporally grounded tasks and precise timestamp annotations. 
In contrast, FTAR~\cite{time-audio} provides temporal annotations, yet is limited to short audio clips and does not cover music.
On the evaluation side, BLAB~\cite{BLAB} provides initial benchmarks for long-form audio but supports only limited tasks such as duration estimation and event localization, primarily on speech data. 
FTAR-test builds upon datasets such as AudioSet~\cite{audioset}, introducing sound signals and supporting tasks including DAC and TAG, but does not cover music and remains limited in modality diversity.
However, it does not cover the TAC task and remains restricted to short audio clips.
Moreover, all existing resources are restricted to English, lacking multilingual coverage.
Overall, the absence of datasets and benchmarks that jointly support long-duration audio, precise temporal annotations, diverse modalities, and multilingual settings remains a key bottleneck for advancing long-form temporal awareness in LALMs.

\begin{table}[thbp]
\centering
\caption{Comparison of long-form audio resources. Abbreviations: Lang. = Language; Max Dur. = Maximum Duration; Sig. = Supported Audio Signals (S = Speech, D = Sound, M = Music); TA = Temporal Annotations.}
\label{tab:comparison_single}
\footnotesize 
\setlength{\tabcolsep}{2pt} 
\begin{tabularx}{\columnwidth}{@{}l l c c c c c @{}}
\toprule
\textbf{} & \textbf{Resource} & \textbf{Lang.} & \textbf{Max Dur.} & \textbf{Sig.} & \textbf{Tasks} & \textbf{TA} \\ \midrule
\multirow{3}{*}{Datasets} & LongAudio-XL & EN & 10min & S, D, M & - & - \\
 & FTAR & EN & 2min & S, D & TAG, DAC & \checkmark \\
 & \textbf{LAT-Chronicle} & \textbf{EN, ZH} & \textbf{30min} & \textbf{S, D, M} & \textbf{TAG, DAC, TAC} & \checkmark \\ \midrule
\multirow{3}{*}{Benchmark} & BLAB & EN & 120min & S & TAG & \checkmark \\
 & FTAR-test & EN & 2min & S, D & TAG, DAC & \checkmark \\
 & \textbf{LAT-Bench} & \textbf{EN, ZH} & \textbf{30min} & \textbf{S, D, M} & \textbf{TAG, DAC, TAC} & \checkmark \\ \bottomrule
\end{tabularx}
\vspace{2pt}
\end{table}

\subsection{Long-form Audio Understanding Methods}
Existing LALMs typically encode audio inputs into embedding sequences via audio encoders, which are then processed by LLMs~\cite{listenthink}. 
The high audio frame rate results in extremely long input sequences, especially in long-form scenarios. 
To handle such inputs, existing approaches mainly adopt two strategies. 
The first extends the context length of LLMs for direct long-context modeling like Qwen3-Omni~\cite{qwen3omni}, Audio Flamingo3~\cite{AF3}, and Gemini series~\cite{gemini-2.5}
But they incur substantial computational and memory costs, and suffer from attention dilution and limited positional encoding extrapolation~\cite{mamba,qwen3-vl}. 
The second adopts sliding-window or chunk-based processing, which reduces computational cost but disrupts global context and breaks temporal continuity. 

\section{LAT-Chronicle}

\begin{figure*}[ht]
  \centering
  \includegraphics[width=1\textwidth]{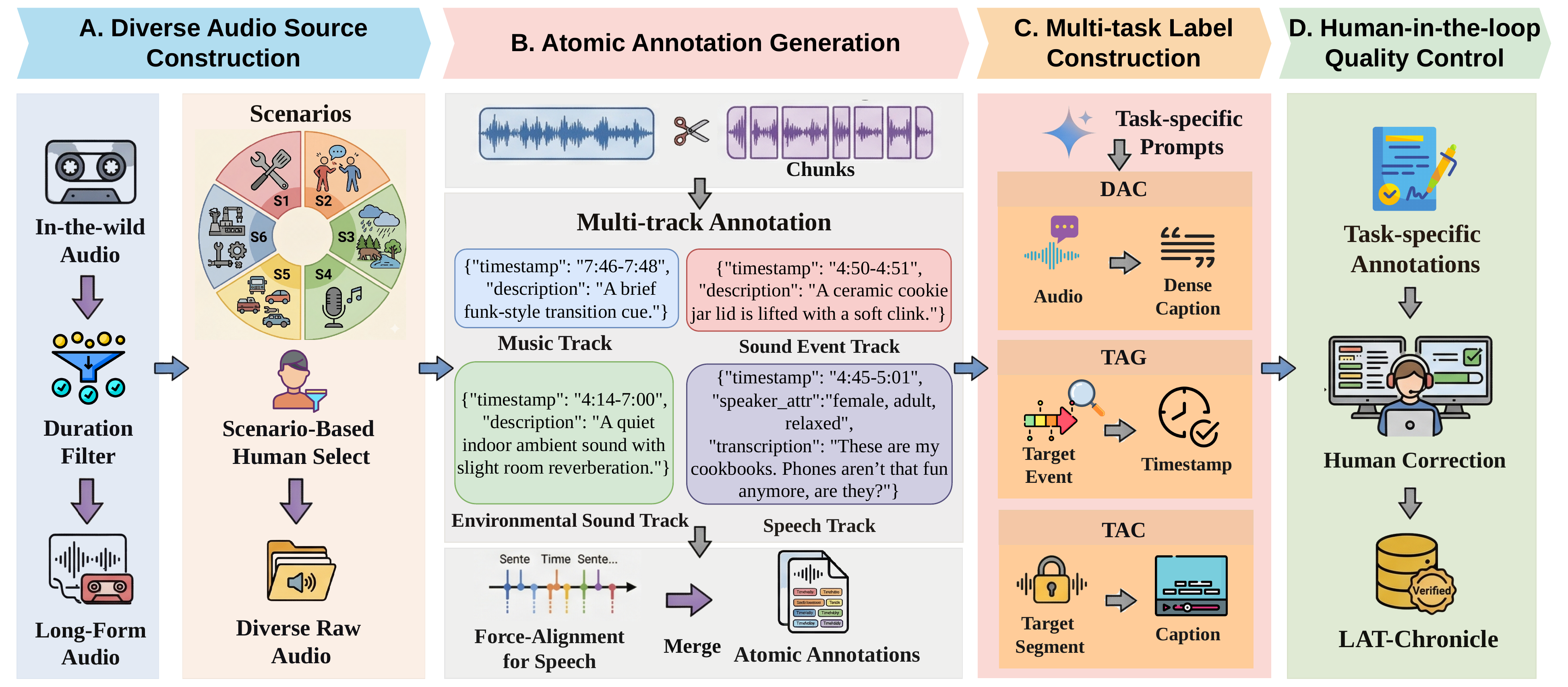}
  \caption{Overview of LAT-Pipe.}
  \label{fig:pipeline}
\end{figure*}

Precise temporal awareness in long-form audio remains underexplored, largely due to a lack of dedicated datasets.
Existing datasets fail to jointly support long-duration audio, fine-grained temporal annotations, and diverse audio modalities. 
To address this gap, we construct LAT-Chronicle to meet these requirements.

\subsection{Task Formulation}
LATA requires models to align audio content with temporal information and perform understanding and reasoning over time. 
To systematically evaluate this ability, we design three complementary tasks over an input audio sequence $A$ with duration $T$: Dense Audio Captioning (DAC), Temporal Audio Grounding (TAG), and Targeted Audio Captioning (TAC), as illustrated in Fig.~\ref{fig:数据示例和错误示例}.

\paragraph{Dense Audio Caption (DAC)}

Given an audio sequence $A$, the goal of DAC is to generate a sequence of temporally localized captions 
$\{(t_s^i, t_e^i, c^i)\}_{i=1}^{N}$,
where $(t_s^i, t_e^i)$ denotes the start and end timestamps of the $i$-th event, and $c^i$ is the corresponding natural language description~\cite{densevedio}. 
DAC requires capturing global audio structure and producing temporally aligned captions with precise timestamps, evaluating both semantic understanding and timestamp accuracy.

\paragraph{Temporal Audio Grounding (TAG)}

Given an audio sequence $A$ and a textual query $q$, TAG aims to localize the corresponding temporal segment $(t_s, t_e)$.
TAG evaluates the ability to precisely ground query-relevant events in time.

\paragraph{Targeted Audio Caption (TAC)}

Given an audio sequence $A$ and a specified temporal segment $(t_s, t_e)$, TAC requires generating a natural language description $c$ for the audio content within the given time interval.
TAC, as the dual of TAG, requires generating a localized, context-aware description for a given temporal interval, assessing the alignment between audio content and temporal information.

\subsection{LAT-Pipe: Data Construction Pipeline}
To support the above tasks with high-quality temporal annotations, we develop LAT-Pipe, a human-in-the-loop pipeline for constructing temporally grounded annotations over long-form audio. 

\renewcommand{\tabularxcolumn}[1]{m{#1}}

\begin{table}[!h]
\centering
\caption{Audio scenario taxonomy in LAT-Pipe.}
\label{tab:scenario_taxonomy}
\footnotesize
\setlength{\tabcolsep}{4pt} 
\begin{tabularx}{\columnwidth}{@{} >{\centering\arraybackslash}m{0.6cm} >{\raggedright\arraybackslash}m{1.5cm} >{\hsize=0.7\hsize\RaggedRight\arraybackslash}X >{\hsize=1.3\hsize\RaggedRight\arraybackslash}X @{} }
\toprule
\textbf{ID} & \textbf{Scenario} & \textbf{Target Focus} & \textbf{Acoustic Characteristics} \\ 
\midrule

S1 & Speech–Sound Interleaving 
& Frequent speech–event alternation  & Strong coupling of speech/sounds (e.g., repairing), frequent transitions. \\ 
\midrule 

S2 & Complex Speech 
& Multi-speaker, emotional variation 
& Fast-paced speech, multiple speakers, and rich emotional dynamics (e.g., debates). \\ 
\midrule

S3 & Noisy Dynamic Environment & Low SNR, background variability & Background noise, sudden acoustic changes, mixed ambient sounds (e.g., Vlogs). \\ 
\midrule

S4 & Speech-Music-Sound Mixed & Foreground speech, background complex music and sound & Speech overlaid with music or original media audio (e.g., reaction videos). \\ 
\midrule

S5 & Clean Structured Speech & Information-dense speech & Clear recording, minimal noise, well-structured monologues or interviews. \\ 
\midrule

S6 & Extremely Complex Audio  & High-Density Speech-Event-Music Overlap & Heavy overlap of speech, music, sound events (e.g., gaming highlights, live streams). \\ 
\bottomrule
\end{tabularx}
\end{table}

\paragraph{A. Diverse Audio Source Construction}
We collect in-the-wild audio data up to 30 minutes. 
To systematically cover real-world conditions, we define a taxonomy of six representative scenarios based on modality composition and acoustic complexity, as shown in Table~\ref{tab:scenario_taxonomy}. 
These scenarios capture diverse combinations of speech, music, and sound events, as well as varying degrees of temporal interleaving, overlap, noise, and information density.
We perform manual selection based on the defined taxonomy, yielding a scenario-balanced dataset

\paragraph{B. Atomic Annotation Generation}
To obtain temporal annotations, we first generate fine-grained atomic annotations by decomposing audio into four parallel tracks: speech, sound events, music, and environmental sound. 
Each track is annotated with temporally aligned information.
For the speech track, we annotate sentence-level timestamps along with transcription and speaker attributes, including gender, age, and emotional state. 
For the remaining tracks, we provide timestamped descriptions.
We segment each audio sample into 5-minute chunks. 
Each chunk is annotated across the four tracks using Gemini-2.5-Pro, one of the most capable models for handling complex audio scenarios and demonstrating strong temporal awareness within short audio segments. 
The chunk-level annotations are then merged to form temporally consistent atomic annotations for the full audio sequence.
To further improve temporal precision in speech, we refine transcription timestamps using a forced alignment model, LLM-ForceAligner~\cite{force-align}, producing sentence-level-aligned timestamps.

\paragraph{C. Multi-task Label Construction}
Based on atomic annotations, we generate task-specific labels for DAC, TAG, and TAC via prompt-driven methods. 
Specifically, we design task-oriented prompts to guide the model in generating dense captions, grounding QA, and targeted descriptions.

\paragraph{D. Human-in-the-loop Quality Control}
To ensure annotation quality, we incorporate human verification at the final stage. Annotated samples are manually reviewed and filtered to eliminate temporal inconsistencies and labeling errors. This human-in-the-loop process significantly improves the dataset's reliability.

\subsection{LAT-Chronicle Statistics}
Based on LAT-Pipe, we construct LAT-Chronicle, a 1.2k hour long-form audio dataset with temporal annotations across six real-world including 1k hours of Chinese data and 200 hours of English data.
We further analyze LAT-Chronicle from three aspects: (1) Duration and Scenario Distribution, (2) Task Coverage and Annotation Statistics, and (3) Quality Analysis.

\begin{figure}[t]
  \centering
  \includegraphics[width=\linewidth]{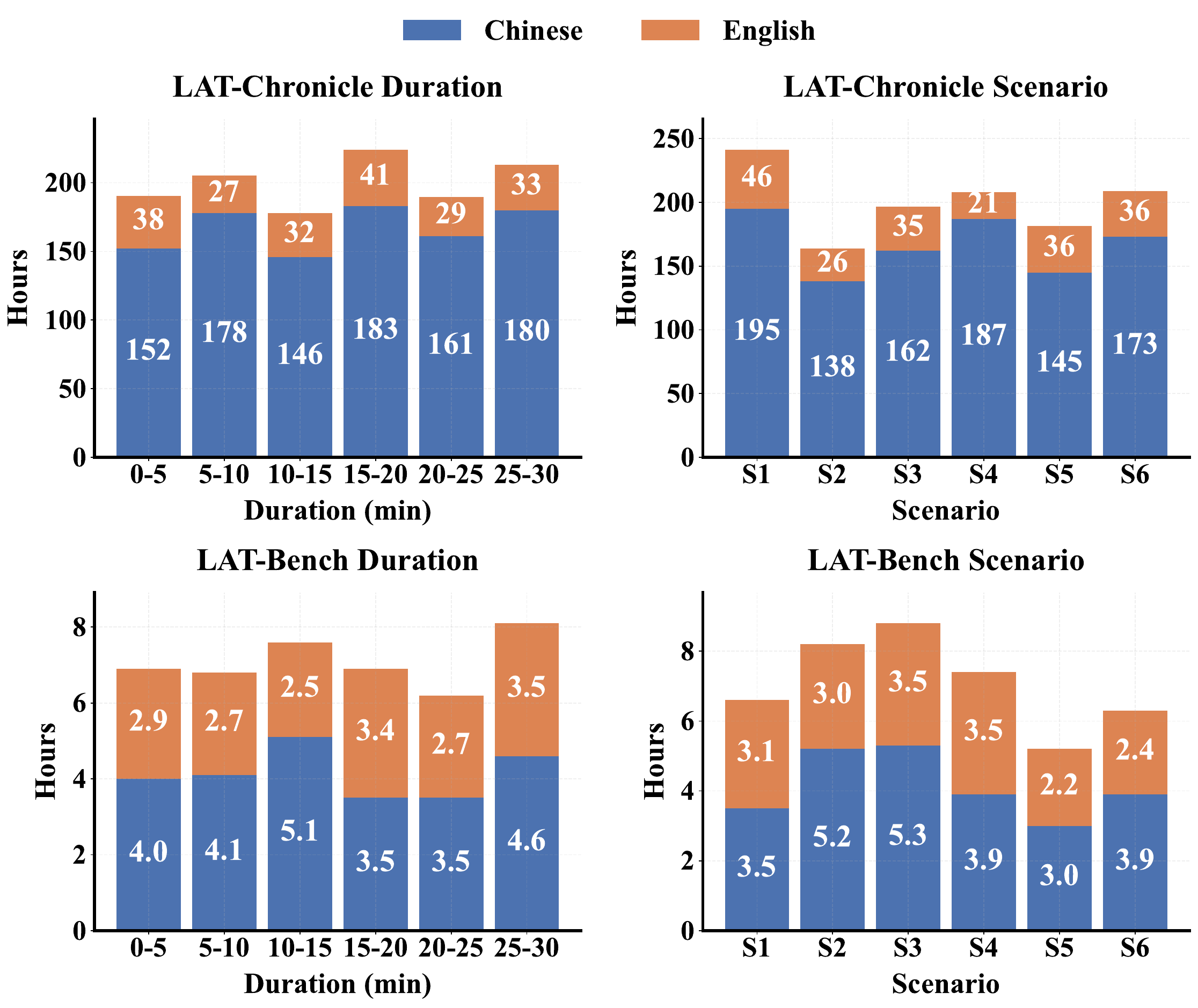}
  \caption{Duration and scenario distributions of LAT-Chronicle and LAT-Bench across Chinese and English.}
  \label{fig:dura}
\end{figure}

\paragraph{Duration and Scenario Distribution.}
We analyze the distribution of LAT-Chronicle across duration ranges, languages, and scenarios. 
Fig.~\ref{fig:dura} shows the data distribution over multiple duration intervals for both Chinese and English, as well as across the six predefined scenarios. 
The results demonstrate balanced coverage across languages, temporal scales, and diverse real-world conditions.
\paragraph{Task Coverage and Annotation Statistics.}
We further examine task distribution and annotation statistics. 
Table~\ref{tab:combined_stats} summarizes the number of samples for each task, along with temporal annotation statistics. 
LAT-Chronicle exhibits high temporal density, with DAC containing a large number of densely annotated events per audio. 
For TAG and TAC, the target intervals are evenly distributed across the beginning, middle, and end of each audio sample. 







\begin{table}[htbp]
\centering
\caption{Temporal annotation statistics and coverage of \textbf{LAT-Chronicle} and \textbf{LAT-Bench}. Abbreviations: Avg. Evt. = Avg. Number of Events; Avg. Evt. Dur. = Avg. Event Duration (s).}
\label{tab:combined_stats}
\footnotesize
\setlength{\tabcolsep}{2pt}

\begin{tabularx}{\columnwidth}{@{} l 
    >{\centering\arraybackslash}m{1.4cm} 
    >{\centering\arraybackslash}m{1.1cm} 
    >{\centering\arraybackslash}m{1.2cm} 
    >{\centering\arraybackslash}X 
    >{\centering\arraybackslash}X 
    >{\centering\arraybackslash}X @{}}
\toprule
\textbf{Task} & \makecell[c]{\textbf{\#Samples}} & \makecell[c]{\textbf{Avg.} \textbf{Evt.}} & \makecell[c]{\textbf{Avg.}\textbf{Evt.}\\ \textbf{Dur (s)}} & \textbf{Start} & \textbf{Middle} & \textbf{End} \\
\midrule
\multicolumn{7}{l}{\textit{\textbf{LAT-Chronicle }}} \\
\midrule
DAC-ZH & 5,537 & 20.18 & 45.51 & -- & -- & -- \\
DAC-EN & 1,459 & 17.32 & 52.93 & -- & -- & -- \\
\addlinespace[2pt]
TAG-ZH & 13,796 & -- & -- & 34.30\% & 32.02\% & 33.68\% \\
TAG-EN & 4,081 & -- & -- & 36.41\% & 33.44\% & 30.15\% \\
\addlinespace[2pt]
TAC-ZH & 12,351 & -- & -- & 33.96\% & 31.72\% & 34.32\% \\
TAC-EN & 4,188 & -- & -- & 32.29\% & 32.60\% & 35.11\% \\

\midrule
\multicolumn{7}{l}{\textit{\textbf{LAT-Bench}}} \\
\midrule
DAC-ZH & 145 & 19.28 & 48.87 & -- & -- & -- \\
DAC-EN & 105 & 18.44 & 50.32 & -- & -- & -- \\
\addlinespace[2pt]
TAG-ZH & 576 & -- & -- & 32.84\% & 34.15\% & 33.01\% \\
TAG-EN & 430 & -- & -- & 31.52\% & 35.67\% & 32.81\% \\
\addlinespace[2pt]
TAC-ZH & 799 & -- & -- & 34.21\% & 32.74\% & 33.05\% \\
TAC-EN & 679 & -- & -- & 33.16\% & 32.48\% & 34.36\% \\
\bottomrule
\end{tabularx}
\end{table}
\paragraph{Annotation Quality Analysis.}
To balance annotation cost and quality, LAT-Chronicle is constructed via a human-in-the-loop pipeline with two levels of quality control. 
At the atomic level, we refine temporal boundaries and reduce annotation errors. 
For speech, forced alignment reduces the average sentence-level timestamp deviation from 672 ms to 102 ms (evaluated on 200 samples). 
For non-speech tracks, we observe a hallucination rate of 1.31\% and an average timestamp deviation of 809 ms. 
At the task level, we apply human verification to ensure annotation consistency. 
TAG and TAC samples are manually reviewed to remove invalid cases, while DAC annotations refine the alignment between temporal boundaries and captions.


\section{LAT-Bench}

LAT-Bench is a human-verified benchmark derived from a held-out subset of LAT-Chronicle with stricter manual annotation, comprising 40 hours of audio, including 25 hours in Chinese and 15 hours in English. 
It supports long-form audio up to 30 minutes and covers three core LATA tasks: DAC, TAG, and TAC. 

\subsection{Benchmark Construction}

LAT-Bench is constructed from a held-out subset of LAT-Chronicle. 
Initial annotations are generated via LAT-Pipe and refined through task-specific human verification.
For TAG and TAC, each sample is reviewed by three annotators, followed by expert verification.
Only validated samples are retained.
Annotation consistency is further evaluated across annotators.
For TAG, agreement is measured by pairwise temporal overlap, achieving an average IoU of 0.897.
For TAC, agreement is measured by the agreement rate, which is 0.895.
For DAC, we adopt a multi-stage consensus process with three annotators.
Annotator A first produces the initial annotation, which is then reviewed by Annotator B. 
Disagreements are resolved through discussion between A and B.
Annotator C conducts an additional round of verification, and ambiguous cases are further discussed with additional annotators when necessary. 
Finally, an expert conducts a final quality control pass.
This process ensures high annotation accuracy and consistency, providing a reliable benchmark for LATA.

\subsection{Benchmark Statistics}

We analyze the composition of LAT-Bench across language, duration, scenario, and task distribution. 
As shown in Fig.~\ref{fig:dura}, durations span up to 30 minutes and remain well-balanced across temporal scales.
All six predefined scenarios are comprehensively covered, reflecting diverse acoustic conditions. 
The temporal positions of TAG and TAC intervals are evenly distributed across the beginning, middle, and end of the audio, as illustrated in Table~\ref {tab:combined_stats}. 








\subsection{Evaluation Metrics}
\label{sec:Evaluation Metrics}

We design task-specific evaluation metrics to assess both temporal alignment and semantic correctness across DAC, TAG, and TAC.

\paragraph{Dense Audio Caption}
Given ground-truth $\{(t_s^i, t_e^i, c^i)\}_{i=1}^{N}$ and predicted outputs $\{(\hat{t}_s^j, \hat{t}_e^j, \hat{c}^j)\}_{j=1}^{M}$, 
We first match each ground-truth segment to the predicted segment with the expected temporal overlap.
The temporal alignment is measured using the Intersection over Union (IoU):
\begin{equation}
\text{IoU}(i, j) = \frac{\min(t_e^i, \hat{t}_e^j) - \max(t_s^i, \hat{t}_s^j)}{\max(t_e^i, \hat{t}_e^j) - \min(t_s^i, \hat{t}_s^j)}.
\end{equation}
A match is considered valid if $\text{IoU}(i, j) \ge \tau$, where $\tau \in \{0.3, 0.5, 0.7\}$. 
For matched pairs, we compute the caption quality using the FENSE score~\cite{fense,midasheng}:
\begin{equation}
s_i = 
\begin{cases}
\text{FENSE}(c^i, \hat{c}^{j^*}), & \text{if } \text{IoU}(i, j^*) \ge \tau \\
0, & \text{otherwise}
\end{cases}
\end{equation}
where $j^* = \arg\max_j \text{IoU}(i, j)$.
The sample-level score is obtained by averaging over all ground-truth segments:
\begin{equation}
S_{\text{DAC}} = \frac{1}{N} \sum_{i=1}^{N} s_i.
\end{equation}
We report results under different IoU thresholds ($\tau = 0.3, 0.5, 0.7$) and use their average as the final DAC score~\cite{activitynet,daceval}.

\paragraph{Temporal Audio Grounding}
Given a ground-truth segment $(t_s, t_e)$ and a predicted segment $(\hat{t}_s, \hat{t}_e)$, 
we evaluate temporal localization using IoU.
We report both the mean IoU (mIoU) and recall under different thresholds ($\tau = 0.3, 0.5, 0.7$), 
where a prediction is considered correct if $\text{IoU} \ge \tau$.

\paragraph{Targeted Audio Caption}
Given a target segment $(t_s, t_e)$, we evaluate the generated caption $\hat{c}$ using the FENSE score:
\begin{equation}
S_{\text{TAC}} = \text{FENSE}(c, \hat{c}).
\end{equation}
This metric measures the semantic quality of captions within the specified temporal region.

\begin{figure*}[!t]
  \centering
\includegraphics[width=1\textwidth]{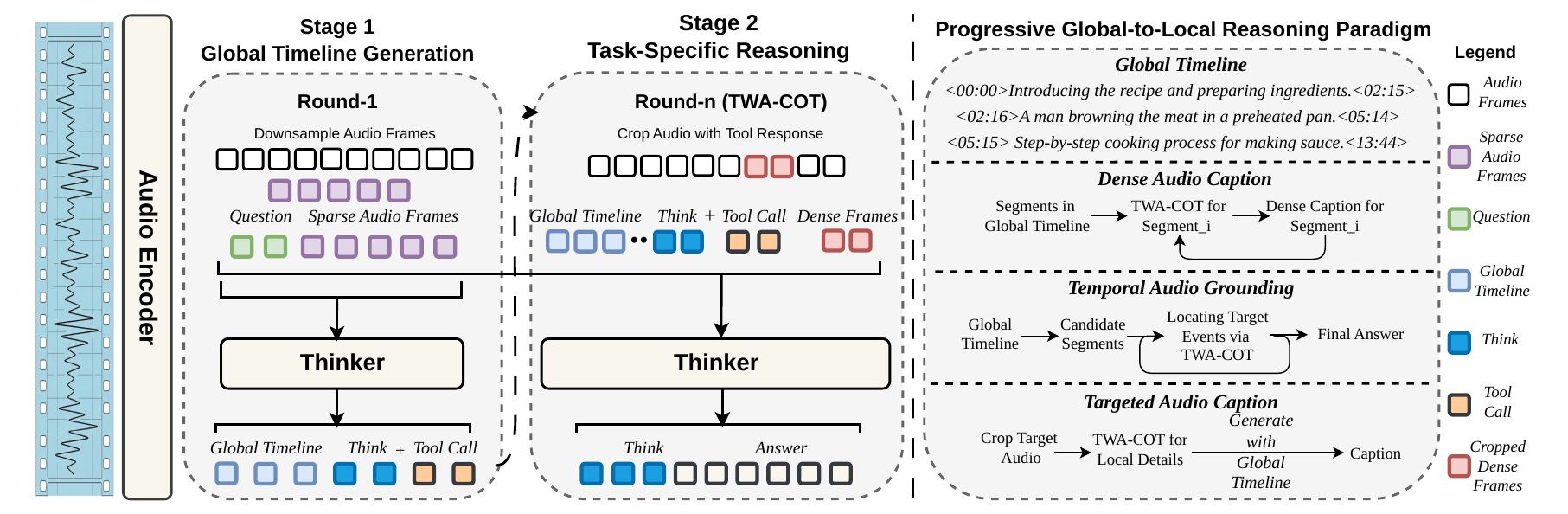}
  \caption{Overall framework of \textbf{LAT-Audio}. 
\textbf{Left:} Long-form audio is downsampled to construct a global timeline for TWA-CoT reasoning.
\textbf{Right:} Progressive global-to-local reasoning paradigm. Temporal-aware tasks are solved by conditioning on the global timeline and iteratively incorporating local audio information via tool use.}
  \label{fig:LAT-Audio}
\end{figure*}

\section{LAT-Audio}
\subsection{Overall Framework}
LAT-Audio formulates LATA as a \textbf{progressive global-to-local reasoning paradigm}. 
It first constructs a global timeline as temporal-semantic context, followed by task-specific reasoning grounded on it, shown as Fig.~\ref{fig:LAT-Audio}).
To realize this paradigm, we introduce \textbf{Think-With-Audio Chain-of-Thought (TWA-CoT)}, a multi-turn reasoning framework with tool use.
It iteratively crops audio segments to incorporate local information for progressive reasoning.

\subsection{Progressive Global-to-Local Reasoning}
The key challenge of LATA is the extremely large temporal range.
To address this, we adopt a divide-and-conquer perspective and propose a novel progressive global-to-local audio reasoning paradigm.
Specifically, we construct a global timeline that structures the entire temporal span into aligned segments.
This global timeline can be formally represented as 
$Z_g = \{(t_s^k, t_e^k, d^k)\}_{k=1}^{K}$, 
where $t_s^k$ and $t_e^k$ denote the start and end timestamps of the $k$-th segment, $d^k$ denotes its corresponding semantic description, and the number of segments \(K\) is set in a small duration-dependent range (e.g., 2–5 for up to 30-minute audio).

The global timeline provides explicit temporal-semantic context to guide subsequent reasoning. 
Given a task, the model first identifies candidate segments relevant to the query and then performs fine-grained reasoning over the selected segments. This progressively narrows the large temporal context, leading to more accurate temporal localization.

\paragraph{Task-specific reasoning}
Building upon the global timeline, we perform task-specific reasoning for different  LATA tasks.
For DAC, the model processes each segment in the global timeline sequentially. 
For each segment, it crops local audio and generates a temporally aligned dense caption conditioned on the global timeline context.
The final output is obtained by concatenating segment-level captions in temporal order.
For TAG, the model first identifies candidate segments from the global timeline by reasoning over the query and coarse temporal-semantic cues.
It then performs fine-grained localization within these candidates.
To obtain detailed information, the model issues tool calls to crop local audio segments and iteratively refines its prediction based on the cropped segments.
The process stops when a final answer is produced or the step limit is reached.
For TAC, the model first crops the target interval to obtain local audio and generates a caption for the segment.
The caption is refined using the global timeline.

\subsection{Think-with-Audio Chain-of-Thought}
To improve temporal accuracy in multi-turn reasoning, we propose TWA-CoT, 
An approach that enhances iterative reasoning by incorporating tool-based audio information.
Each TWA-CoT iteration consists of three steps: 

(1) \textbf{Think}: the model performs deliberation based on the task and current reasoning state.
It then decides the next action: whether to invoke a tool call to collect further information or to produce the final answer.

(2) \textbf{Tool call}: the model invokes \texttt{crop\_audio} with a predicted start and end time to extract a local clip from the original audio; 

(3) \textbf{Tool response}: the model obtains the cropped audio.

Formally, TWA-CoT is defined as:
\begin{equation}
r_{i+1} = \mathcal{T}(r_i, A_i, Z_g), \quad 
A_i = \texttt{crop\_audio}(A, \tilde{t}_s^{\,i}, \tilde{t}_e^{\,i}),
\end{equation}
where $Z_g$ is the global timeline and $A_i$ is the cropped audio, the crop range $(\tilde{t}_s^{\,i}, \tilde{t}_e^{\,i})$ is predicted from the current reasoning state.
In contrast, standard CoT updates reasoning purely in the textual cue~\cite{think-with-iamge,longvt},
TWA-CoT incorporates audio information at each step for iterative verification and correction.
Reasoning terminates either when the model outputs a final answer that satisfies the task-specific format, or when the maximum number of iterations is reached. 
In our implementation, the maximum number of reasoning steps is set to four.

\subsection{Model Architecture}
We adopt Qwen3-Omni as the backbone. 
Given a prompt and a long audio clip, the input is encoded and temporally downsampled.
The thinker-LLM then generates a global timeline.
This global timeline, together with the query and audio features, forms the context for subsequent reasoning.
The model then performs multi-turn reasoning via TWA-CoT, with tool-based audio cropping at each step.
Cropped audio is encoded without downsampling and used for subsequent reasoning.

\paragraph{On-Demand Sampling}
To reduce computation cost, we adopt sparse sampling for long audio~\cite{visionzip}. 
Specifically, we apply $2\times$ temporal downsampling when generating the global timeline.
During reasoning, we use full-resolution audio frames to preserve sufficient local detail. 
This strategy reduces context length, alleviating attention dilution and positional extrapolation, while lowering the cost of multi-turn reasoning.
In practice, it reduces input tokens to approximately half for TAG/TAC and to around $1.5\times$ for DAC.

\subsection{Training Strategy}
\paragraph{Stage 1: Global Timeline Generation SFT}
In the first stage, we train the model to generate global timeline via supervised fine-tuning (SFT). 
The global timeline annotations are generated by an LLM based on atomic annotations.

\paragraph{Stage 2: Full-Trajectory SFT}
In the second stage, we train the model to perform full task-specific reasoning trajectories. 
We construct training trajectories using an LLM with oracle access, which generates reasoning traces conditioned on task QA, the global timeline, and atomic annotations.

\paragraph{Stage 3: Reinforcement Learning}
In the final stage, we apply reinforcement learning (RL) with Group Relative Policy Optimization (GRPO)~\cite{grpo} to further improve reasoning quality.
Training data is constructed by sampling each instance 8 times using the Stage 2 model and selecting trajectories that include both correct and incorrect reasoning as supervision signals.
This improves temporal reasoning robustness.

Both global timeline generation and TWA-CoT reasoning follow predefined structured output schemas. The timeline is generated as an ordered list of temporal-semantic segments, while each reasoning trajectory follows a fixed Think–Tool Call–Tool Response–Answer pattern. 
\paragraph{Reward Design.}

The total reward is computed as the sum of format rewards and task rewards.

Let $K$ be the number of sampled rollouts for each input, and let $y^{(k)}$ denote the full output trajectory of the $k$-th rollout, including reasoning steps and the final prediction $\hat{a}^{(k)}$.

\textbf{Format Reward.}
Let $S$ denote the predefined output schema, which specifies valid reasoning structure. 
The format reward is defined as:
\begin{equation}
R_{\text{format}}^{(k)} =
\begin{cases}
1, & \text{if } y^{(k)} \text{ follows } S, \\
0, & \text{otherwise}.
\end{cases}
\end{equation}

\textbf{Task Reward.}
The task reward $R_{\text{task}}^{(k)}$ is defined based on task-specific evaluation metrics.

For TAG, let $\hat{z}^{(k)}$ denote the predicted interval and $z^\star$ the ground-truth interval. 
The reward is defined as:
\begin{equation}
R_{\text{task}}^{(k)} = \text{IoU}\big(\hat{z}^{(k)}, z^\star\big) + \frac{1}{N} \sum_{i=1}^{N} \mathbb{I}\big( |c_i^{(k)} - c^\star| < |c_{i-1}^{(k)} - c^\star| \big),
\label{eq:reward_task}
\end{equation}
where $N$ is the number of steps, $c_i^{(k)}$ is the midpoint of the predicted segment at step $i$, and $c^\star$ is the midpoint of the ground-truth interval.

For DAC, we directly use the evaluation score as the reward:
\begin{equation}
R_{\text{task}}^{(k)} = S_{\text{DAC}}^{(k)},
\end{equation}
where $S_{\text{DAC}}^{(k)}$ is the dense captioning score defined in Sec.~\ref{sec:Evaluation Metrics}.

For TAC, the reward is defined based on caption quality:
\begin{equation}
R_{\text{task}}^{(k)} = \text{FENSE}(\hat{cap}^{(k)}, cap^\star),
\end{equation}
where $\hat{cap}^{(k)}$ is the generated caption and $cap^\star$ is the reference.





\section{Experiments}
\subsection{Experimental Setup}
\paragraph{Implementation Details.}
We implement LAT-Audio based on the Qwen3-Omni-30B-A3B-Instruct~\cite{qwen3omni} using the Swift framework~\cite{swift}. 
Audio features are temporally downsampled by $2\times$. 
We use full-parameter fine-tuning with learning rates of $1\mathrm{e}{-6}$, $1\mathrm{e}{-5}$, and $1\mathrm{e}{-6}$ for Stages 1–3, respectively, and set the GRPO group size to 8.
\paragraph{Training Data.}
All training data are derived from LAT-Chronicle, including global timelines and full reasoning trajectories. 
In Stage 1, we generate global timelines for each audio as supervision, resulting in 7K training samples. 
In Stage 2, we construct full CoT trajectories for each QA pair, yielding 30K samples.
In Stage 3, we perform multiple rounds of sampling to select high-quality trajectories with balanced numbers of correct and incorrect samples, resulting in 2.5K training instances. 

\paragraph{Baselines.}

\begin{table*}[t]
\centering
\caption{Main results on LAT-Bench and BLAB. For LAT-Bench, results are reported as \textbf{Chinese / English}. For BLAB, results are reported on a subset of audio samples with durations up to 30 minutes. \textbf{Bold} and \underline{underline} indicate the best and second-best performance in each column, respectively. \textbf{Recall@$\tau$} and \textbf{score@$\tau$} denote the Recall and DAC score at an IoU threshold of $\tau$.}
\label{tab:main_results}
\footnotesize 
\setlength{\tabcolsep}{1.8pt} 

\begin{tabular}{l cccc cccc c cccc}
\toprule
\multirow{3}{*}{\textbf{Model}} & \multicolumn{9}{c}{\textbf{LAT-Bench}} & \multicolumn{4}{c}{\textbf{BLAB}} \\
\cmidrule(lr){2-10} \cmidrule(lr){11-14} 
& \multicolumn{4}{c}{TAG} & \multicolumn{4}{c}{DAC} & TAC & \multicolumn{4}{c}{Advertisement Localization} \\
\cmidrule(lr){2-5} \cmidrule(lr){6-9} \cmidrule(lr){10-10} \cmidrule(lr){11-14}
& mIoU & Recall@0.3 & Recall@0.5 & Recall@0.7 & Avg\_score & Score@0.3 & Score@0.5 & Score@0.7 & Fense & mIoU & Recall@0.3 & Recall@0.5 & Recall@0.7 \\
\midrule
\multicolumn{14}{l}{\textit{Main Models}} \\
\textbf{LAT-Audio (Ours)} & \textbf{47.2/50.0} & \textbf{63.7/68.1} & \textbf{49.0/54.1} & \textbf{32.6/34.5} & \textbf{46.8/48.6} & \textbf{61.0}/\underline{61.4} & \textbf{45.5/49.5} & \textbf{33.7/34.8} & \textbf{62.0/68.7} & \textbf{49.3} & \textbf{66.7} & \underline{51.4} & \textbf{30.1} \\
Gemini-2.5-Pro~\cite{gemini-2.5} & \underline{40.3/45.3} & \underline{61.3/65.2} & \underline{48.7/53.9} & \underline{26.1/27.7} & 41.8/42.8 & \underline{60.4}/61.1 & 41.9/45.3 & 23.1/21.9 & \underline{58.1}/63.0 & \underline{43.8} & \underline{64.4} & \textbf{55.6} & \underline{29.0} \\
Gemini-3.0-Pro~\cite{gemini-2.5} & 34.6/41.0 & 50.9/51.4 & 32.8/44.8 & 22.8/22.9 & \underline{42.5/46.2} & 59.6/\textbf{61.9} & \underline{43.1/46.0} & \underline{24.9/30.8} & 57.1/\underline{63.2} & 36.2 & 53.2 & 36.8 & 23.2 \\
Qwen3-Omni~\cite{qwen3omni} & 14.8/15.8 & 21.4/26.4 & 12.4/16.0 & 7.0/7.0 & 9.1/10.4 & 16.4/17.7 & 6.5/8.0 & 4.3/5.7 & 28.4/31.0 & 15.7 & 22.4 & 16.3 & 9.6 \\
\midrule
\multicolumn{14}{l}{\textit{Sliding Window (SW)}} \\
Audio-Flamingo3-SW~\cite{AF3} & 3.7/4.1 & 6.4/7.0 & 3.9/4.3 & 2.9/2.9 & 2.2/2.7 & 3.2/4.4 & 2.2/2.5 & 1.0/1.1 & 45.3/51.6 & 5.0 & 8.1 & 5.7 & 3.3 \\
Qwen3-Omni-SW~\cite{qwen3omni} & 22.8/26.2 & 37.1/41.9 & 22.2/25.8 & 14.4/15.5 & 8.9/10.6 & 15.8/18.8 & 7.6/8.5 & 3.3/4.4 & 51.5/53.7 & 26.3 & 36.7 & 29.4 & 20.2 \\
Step-Audio-R1.1-SW~\cite{step-audio-r1} & 8.1/9.0 & 10.9/11.2 & 8.5/9.3 & 5.1/6.1 & 3.4/4.1 & 4.2/5.2 & 3.9/4.3 & 2.1/2.9 & 48.9/51.2 & 6.1 & 8.3 & 6.0 & 5.6 \\
Gemini-2.5-Pro-SW~\cite{gemini-2.5} & 35.8/40.6 & 49.2/54.0 & 36.1/42.7 & 23.3/29.7 & 38.8/40.4 & 48.7/55.1 & 39.9/43.3 & 27.8/31.9 & 52.4/58.1 & 34.9 & 45.7 & 32.8 & 20.3 \\
Time-Audio-SW~\cite{time-audio} & --/2.5 & --/2.9 & --/2.3 & --/1.1 & --/1.6 & --/2.3 & --/1.5 & --/1.0 & --/35.6 & 3.8 & 4.2 & 3.9 & 3.0 \\
\midrule
\multicolumn{14}{l}{\textit{Ablation Study}} \\
QA-only SFT & 36.4/39.2 & 49.7/54.2 & 38.6/41.7 & 21.7/24.5 & 37.5/40.2 & 52.3/55.6 & 38.3/41.8 & 22.0/23.1 & 52.3/59.0 & -- & -- & -- & -- \\
w/o Global Timeline & 41.6/45.3 & 45.9/52.8 & 42.8/48.4 & 28.1/29.5 & 42.3/46.0 & 53.7/59.9 & 44.9/49.0 & 28.3/29.1 & 58.8/66.1 & -- & -- & -- & -- \\
w/o TWA-CoT & 38.9/40.3 & 51.3/55.0 & 40.1/43.8 & 24.9/26.7 & 39.6/41.9 & 54.1/57.3 & 40.9/42.8 & 23.9/25.6 & 53.6/60.8 & -- & -- & -- & -- \\
w/o Stage3-RL & 45.3/47.3 & 60.8/68.0 & 45.1/50.3 & 30.2/31.1 & 44.1/46.2 & 59.4/60.3 & 41.4/46.0 & 31.7/32.2 & 60.2/65.5 & -- & -- & -- & -- \\
w/o Stage1-SFT+Stage3-RL & 42.3/45.2 & 54.7/60.3 & 43.5/52.9 & 29.0/27.9 & 39.1/39.9 & 50.9/51.7 & 38.4/40.6 & 27.8/27.3 & 56.5/60.1 & -- & -- & -- & -- \\

Downsampling $\times 1$ & 45.4/48.7 & 61.0/64.8 & 48.8/52.0 & 20.3/30.1 & 43.2/47.3 & 55.6/61.9 & 42.2/47.9 & 31.7/32.0 & 60.3/66.6 & -- & -- & -- & --  \\
Downsampling $\times 4$ & 39.1/41.5 & 53.7/61.9 & 41.1/41.6 & 27.3/24.2 & 40.9/43.1 & 52.9/56.1 & 40.1/43.9 & 29.8/29.4 & 58.6/65.5 & -- & -- & -- & --  \\
Downsampling $\times 8$ & 27.1/27.8 & 39.4/45.2 & 26.6/29.8 & 15.8/15.7 & 27.6/32.9 & 37.9/45.7 & 26.8/32.9 & 18.1/20.1 & 57.8/62.6 & -- & -- & -- & --  \\
\bottomrule
\end{tabular}
\end{table*}

We consider two categories of baselines:

(1) End-to-end long-context LALMs, which directly process up to 30-minute audio.

(2) Sliding-window methods, where audio is split into 1-minute chunks for task-specific inference. 
For DAC, dense captions are generated per chunk;
For TAG, chunks are processed sequentially to determine whether they contain the target event. 
Once a chunk is predicted as positive, the model outputs the corresponding timestamp as the final answer; otherwise, it proceeds to the next chunk.;
For TAC, captions are generated directly from the target segment.

\paragraph{Evaluation Benchmarks.}
We evaluate on both LAT-Bench and BLAB.
LAT-Bench measures performance on DAC, TAG, and TAC using task-specific metrics.
For BLAB, we adopt a transferable task:
Advertisement localization, evaluated using BLAB-TAG metrics.
We restrict BLAB to audio durations of up to 30 minutes.

\subsection{Main Results}
Table~\ref{tab:main_results} presents the main results on LAT-Bench and BLAB. 
LAT-Audio surpasses prior methods across all tasks. 
For temporal localization, LAT-Audio achieves significant improvements over strong baselines, outperforming Gemini-2.5-Pro by 17.1\% on LAT-Bench-TAG and 13.8\% on BLAB advertisement localization.
For TAC, LAT-Audio attains the highest FENSE score, demonstrating accurate alignment between temporal segments and semantic content. 
For DAC, LAT-Audio surpasses Gemini-3.0-Pro, achieving a relative improvement of 10.11\% in average score, demonstrating its advantage in dense temporal-semantic understanding.

\paragraph{Analysis of Sliding-Window Methods.}
We observe that sliding-window methods exhibit inconsistent behavior across models. 
For strong long-context models such as Gemini-2.5-Pro, the sliding-window variant leads to substantial performance degradation, indicating that breaking global context and temporal continuity harms temporal reasoning. 
In contrast, Qwen3-Omni benefits from the sliding-window strategy. 
This is because its LATA under long-context settings is limited, while it performs relatively better on short segments. 
In this case, the gain from reduced context length outweighs the loss of global information.
For models such as Audio-Flamingo3 and Step-Audio-R1.1, which lack temporal awareness even on short audio segments, sliding-window processing does not improve performance. 
Similarly, Time-Audio shows poor generalization on LAT-Bench due to limited training data, resulting in overall weak performance.
Overall, sliding-window methods degrade performance for models with strong long-form temporal reasoning by destroying global structure, while providing limited gains for models that operate effectively only on short contexts and little benefit for models without temporal awareness.

\begin{figure}[t]
  \centering
  \includegraphics[width=0.98\linewidth]{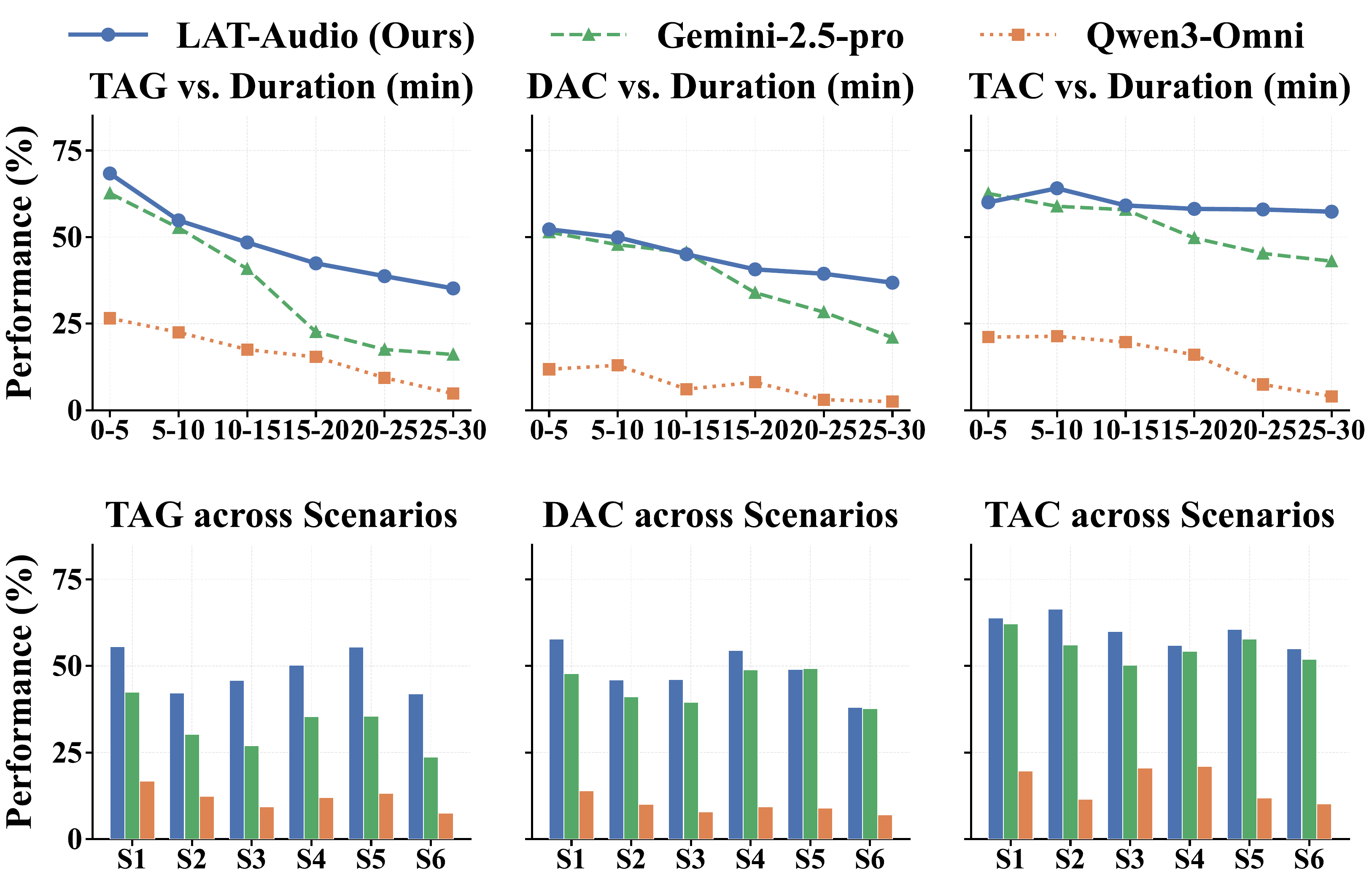}
  \caption{Model performance across duration and scenarios.}
  \label{fig:rubost}
\end{figure}

\subsection{Ablation Study}
Table~\ref{tab:main_results} presents ablation results on key components, training strategy, and temporal downsampling.
QA-only SFT, which fine-tunes Qwen3-Omni using only the original QA pairs, improves over the base Qwen3-Omni, validating the effectiveness of LAT-Chronicle.
However, it remains nearly 22\% below LAT-Audio, highlighting the importance of proposed progressive global-to-local reasoning paradigm.
Removing the global timeline (w/o Global Timeline) results in a clear drop across all tasks, underscoring the importance of explicit temporal structuring. 
Similarly, removing TWA-CoT (w/o TWA-CoT) degrades performance, confirming the necessity of iterative, evidence-grounded reasoning. 
Combining both yields the best results, indicating that global context and local refinement are complementary.
Removing RL training (w/o Stage3-RL) results in consistent degradation, suggesting that RL helps refine multi-turn decision-making.
Further removing global timeline training (w/o Stage1-SFT+Stage3-RL) leads to additional drops, confirming that global-timeline-SFT is essential.
For temporal downsampling, a 2$\times$ downsampling yields around 5\% performance gain by reducing context length. 
However, more aggressive downsampling like 4$\times$ and 8$\times$ leads to significant performance degradation.


\subsection{Robustness Analysis}

Fig.~\ref{fig:rubost} presents the performance of LAT-Audio, Gemini-2.5-Pro, and Qwen3-Omni across different audio durations and scenarios. 
Gemini-2.5-Pro shows a sharp decline beyond 15 minutes (e.g., TAG drops from 62.6 to 16.1), while Qwen3-Omni degrades steadily with duration. 
In contrast, LAT-Audio exhibits a much smaller drop (68.4 to 35.2), demonstrating stronger robustness to long-form audio.
Across scenarios, temporal awareness varies significantly. 
In particular, all models experience notable performance drops in S6 (complex acoustic environments such as live streams), indicating that overlapping and high-density audio content make temporal reasoning more difficult.


\section{Conclusion and Future Work}

In this work, we study LATA and identify the large temporal context as the core challenge, making accurate temporal alignment difficult. 
To address this, we introduce LAT-Chronicle and LAT-Bench for training and evaluation, and propose LAT-Audio, a global-to-local reasoning framework with TWA-CoT for iterative audio-grounded reasoning. 
By structuring temporal reasoning and incorporating TWA-CoT, our approach enables more accurate and stable temporal alignment.
Experiments show that LAT-Audio surpasses existing methods and improves robustness as audio duration increases. 
Despite these results, our framework has several limitations.
First, multi-turn reasoning with tool use introduces additional computational overhead, limiting efficiency in real-time scenarios.
Second, the framework focuses on single-audio inputs and does not fully extend to more complex multimodal settings.
In future work, we plan to improve the efficiency of long-form temporal reasoning and extend the framework to broader multimodal scenarios, such as audio-visual understanding.

\newpage
\onecolumn
\section{Appendix}
\subsection{Atomic Annotation Examples}
We provide a representative example of atomic annotation (truncated to the first 1 minutes for brevity).

\begin{tcolorbox}[
colback=white,
colframe=black,
title=Atomic Annotation Example ,
breakable,
fonttitle=\bfseries,
sharp corners,
left=3mm,
right=3mm,
width=\textwidth
]

\begin{lstlisting}
{"summary": "A male vlogger explores Chinese electric cars in a showroom, expressing excitement about their advanced features and surprisingly low prices.",
  "duration": "01:00",
  "tracks": {
    "speech": [
      {"start": "00:03", "end": "00:09",
        "speaker_attr": "Male, Young Adult, Excited",
        "content": "The speaker is surprised by low price and highlights dynamic interior lighting synced with music",
        "transcription": "Is 29,000 US for this car. Look at the lights moving with the music. Very nice."},
      {"start": "00:10", "end": "00:15",
        "speaker_attr": "Male, Young Adult, Amazed",
        "content": "He reacts to futuristic features, noting that cameras replace traditional side mirrors.",
        "transcription": "Whoa, what's going on? Look at that. This is the future. You don't need side mirrors anymore."},
      {"start": "00:15", "end": "00:21",
        "speaker_attr": "Male, Young Adult, Impressed",
        "content": "He admires the car design and large display, suggesting it supports media functions.",
        "transcription": "Look at this ride. My god. Could you pick a movie from here?"},
      {"start": "00:22", "end": "00:27",
        "speaker_attr": "Male, Young Adult, Impressed",
        "content": "He considers it one of the best SUVs and notices interior-facing cameras.",
        "transcription": "This has to be the nicest SUV we've seen. There's even a camera inside."},
      {"start": "00:27", "end": "00:36",
        "speaker_attr": "Male, Young Adult, Shocked",
        "content": "He is amazed by voice-controlled features like opening the sunroof and compares pricing globally.",
        "transcription": "Could you get this for $41,000 anywhere?"},
      {"start": "00:39", "end": "00:55",
        "speaker_attr": "Male, Young Adult, Excited",
        "content": "He praises build quality, interior features, and highlights seat massage functions.",
        "transcription": "Look at the quality. Even the back seats massage you."}],
    "events": [
      {"start": "00:32", "end": "00:33",
        "description": "Soft electronic beep from car interaction."}],
    "music": [
      {"start": "00:00", "end": "00:09",
        "description": "Light electronic background music creating a modern showroom atmosphere."}],
    "background": [
      {"start": "00:00", "end": "01:00",
        "description": "A quiet indoor showroom environment with minimal ambient noise."}]
  }
}
\end{lstlisting}

\end{tcolorbox}

\subsection{Reasoning Trajectories for DAC, TAC, and TAG}
We present representative multi-turn reasoning trajectories for DAC, TAC, and TAG, illustrating how global timeline guidance and iterative audio evidence retrieval enable precise and consistent temporal reasoning in long-form audio.
The following trajectory demonstrates how the model decomposes long-form audio into structured segments under global timeline guidance, while iteratively generating local dense captions.
\begin{tcolorbox}[
colback=white,
colframe=black,
title=Reasoning Trajectory Examples,
breakable,
fonttitle=\bfseries,
sharp corners,
left=3mm,
right=3mm,
width=\textwidth
]

\textbf{Dense Audio Caption}

\begin{lstlisting}
{"messages": [
      {"role": "user",
        "content": "<audio> Provide a dense audio caption for this audio segment."},
      {"role": "assistant",
        "content": "<global_timeline>\n
        [00:00 - 02:10] The speaker opens with a nostalgic anecdote about 'Good Will Hunting'...\n
        [02:11 - 04:10] Adopting a sincere tone, the speaker introduces 'Simulation Theory'...\n
        [04:11 - 05:52] Transitioning from satire to inspiration, the speaker delivers a call to action...\n
        </global_timeline>\n
        <think>According to the global timeline, observing the interval [00:00 - 02:10].</think>"},
      {"role": "assistant",
        "tool_call": {"name": "crop_audio","arguments": {"start_sec": 0.0, "end_sec": 130.0}}},
      {"role": "tool_response",
        "content": "Segment extracted: <audio>"},
      {"role": "assistant",
        "content": "<think>This segment contains 2 key description points.</think>\n
        [00:00 - 01:13]: The speaker recalls growing up near MIT and shares an anecdote about a fake equation inspiring Good Will Hunting....\n
        [01:14 - 02:10]: He humorously reads a satirical review of the film, triggering audience laughter...."},
      {"role": "assistant",
        "content": "<think>Continue to observe the next interval.</think>"},
      {"role": "assistant",
        "tool_call": {"name": "crop_audio", "arguments": {"start_sec": 131.0, "end_sec": 250.0}}},
      {"role": "tool_response", "content": "Segment extracted: <audio>"},
      {"role": "assistant",
        "content": "<think>This segment contains 2 key description points.</think>\n
        [02:11 - 02:46]: The speaker introduces Simulation Theory....\n
        [02:47 - 04:10]: He explains the concept and delivers political satire, triggering applause...."},
      {"role": "assistant","content": "<think>Continue to observe the next interval.</think>"},
      {"role": "assistant",
        "tool_call": {"name": "crop_audio","arguments": {"start_sec": 251.0, "end_sec": 352.0}}},
      {"role": "tool_response", "content": "Segment extracted: <audio>"},
      {"role": "assistant",
        "content": "<think>This segment contains 1 key description point.</think>\n
        [04:11 - 05:52]: The speaker delivers an inspiring call to action on global challenges...."}]
  }
\end{lstlisting}

\vspace{2mm}
\hrule
\vspace{2mm}

\textbf{Targeted Audio Caption}

\begin{lstlisting}
{"messages": [
      {"role": "user", "content": "<audio> Please listen carefully to [01:49 - 02:10]. Combine speech, background music, events, and environment to produce a comprehensive description."},
      {"role": "assistant",
        "content": "<global_timeline>\n
        [00:00 - 02:26] In a lively studio, a host introduces a charity game...\n
        [02:26 - 03:43] The final round concludes with a donation and applause...\n
        </global_timeline>\n
        <think>This interval belongs to the first stage, reinforcing the comedic tone.</think>"},
      {"role": "assistant",
        "tool_call": {"name": "crop_audio", "arguments": {"start_sec": 109.0, "end_sec": 130.0}}},
      {"role": "tool_response", "content": "Segment extracted: <audio>"},
      {"role": "assistant",
        "content": "<think>A female guest speaks humorously about a bowl she \"stole\", with audience laughter confirming the comedic tone.</think>\n
        Against a background of live audience murmurs, the guest humorously explains that her house is full of things stolen from her mother. She presents a bowl used for AirPods and jokes it might be a Picasso piece, triggering loud audience laughter."}]
  }
\end{lstlisting}

\vspace{2mm}
\hrule
\vspace{2mm}

\textbf{Temporal Audio Grounding}

\begin{lstlisting}
{"messages": [
      {"role": "user",
        "content": "<audio> Please locate the final conclusion of the interview where the speaker, in a calm and firm tone, states that he should return to his original intention and focus on \"playing the piano well\". Output strictly in [MM:SS - MM:SS]."},
      {"role": "assistant",
        "content": "<global_timeline>\n
        [00:00 - 01:27] Lang Lang reflects on competition and growth...\n
        [01:27 - 03:30] He demonstrates music as a \"gateway\" to the world...\n
        [03:33 - 08:51] He reflects on artistic identity and concludes by returning to his core belief of focusing on piano.\n
        </global_timeline>\n
        <think>The final conclusion must lie in the last segment [03:33 - 08:51]. I first probe near the end: [07:51 - 08:11].</think>"},
      {"role": "assistant", "tool_call": {"name": "crop_audio", "arguments": {"start_sec": 471.0, "end_sec": 491.0}}},
      {"role": "tool_response", "content": "Segment extracted: <audio>"},
      {"role": "assistant",
        "content": "<think>This segment reflects emotional retrospection rather than the final answer. The true conclusion should follow, so I move closer to the end.</think>\n
        [08:42 - 08:51]"}]
  }
\end{lstlisting}

\end{tcolorbox}

\subsection{Baseline Evaluation Details}
\label{sec:baseline setup}
\paragraph{End-to-End Models}

We evaluate several representative end-to-end Large Audio Language Models (LALMs), including Gemini-2.5-Pro, Gemini-3.0-Pro, and Qwen3-Omni.

For each model, we conduct three independent runs and report the averaged performance to reduce randomness.

For Gemini-series models, we access the models via the Google AI Studio API with default decoding parameters. 
For Qwen3-Omni, we deploy the model locally using vLLM for inference.

\paragraph{Prompting Strategy.}
For TAG and TAC tasks, we directly use the original task queries as prompts to preserve their semantic specificity and avoid introducing additional bias.

For DAC, to ensure consistent task interpretation across models, we adopt a unified instruction prompt based on our DAC annotation protocol. This reduces ambiguity and enforces comparable output structures across different models. The detailed prompt is shown in Box~\ref{box:dac_prompt}.

\begin{tcolorbox}[
colback=white,
colframe=black,
title=DAC Instruction Prompt for Baseline Models,
label={box:dac_prompt},
breakable,
fonttitle=\bfseries,
sharp corners,
left=3mm,
right=3mm,
width=\textwidth
]

\textbf{Role}

You are an expert in \textbf{Dense Audio Caption} with strong auditory perception and narrative understanding abilities. 
Your goal is to transform continuous audio signals into \textbf{semantically coherent, detail-rich, and temporally structured dense caption}.

\textbf{Task}

Perform \textbf{Dense Audio Captioning} on the input audio. 
You must jointly analyze speech content, paralinguistic features (e.g., tone and emotion), and acoustic environment (e.g., background, events and music), and produce a high-density descriptive timeline.

\textbf{Core Guidelines}

\textbf{1. Semantic Primacy}

\begin{itemize}
\item Avoid trivial descriptions such as ``someone is speaking''.
\item Capture the \textbf{core intent, topic, or narrative content} of speech.
\item Explicitly mention key entities such as names, locations, or domain-specific terms when present.
\item Infer relationships between speakers when possible (e.g., interviewer–interviewee, argument, hierarchy).
\end{itemize}

\textbf{2. Integrated Narrative Style}

\begin{itemize}
\item Do not list acoustic elements mechanically (e.g., ``there is background music'').
\item Integrate environmental sounds into the narrative:
\begin{itemize}
\item Bad: A man is speaking, background is noisy.
\item Good: Amid loud construction noise and machinery, a man raises his voice to deliver instructions.
\end{itemize}
\item Make emotions explicit (e.g., anxious, sarcastic, hesitant).
\end{itemize}

\textbf{3. Adaptive Segmentation}

\begin{itemize}
\item Do not segment based on short pauses (1--2 seconds).
\item Segment only when:
\begin{enumerate}
\item Topic or event changes
\item Acoustic scene changes significantly
\item Emotional state shifts abruptly
\end{enumerate}
\item Maintain continuity within the same scene.
\end{itemize}

\textbf{4. Event Sensitivity}

\begin{itemize}
\item Capture short but meaningful events (e.g., door slam, glass breaking, siren). These events must be included as they often drive the narrative.
\end{itemize}

\textbf{Output Format}

\begin{lstlisting}
[{"start": "MM:SS", "end": "MM:SS", "caption": "Detailed narrative description integrating semantics and acoustic context."}]
\end{lstlisting}

\end{tcolorbox}

\paragraph{Sliding-Window Baseline}

To further evaluate the impact of global context, we construct a sliding-window baseline for all tasks. 
Specifically, long-form audio is divided into non-overlapping chunks of 60 seconds, which provides a balance between sufficient local context for semantic understanding and compatibility with the input length constraints of most models.

For DAC, each chunk is fed into the model using the same prompt as in Box~\ref{box:dac_prompt} to generate dense captions. 
The predicted timestamps are defined within each chunk, and are subsequently converted to global timestamps by offsetting with the corresponding chunk start time. 
All chunk-level predictions are then concatenated as the final output.

For TAC, we directly crop the target interval from the audio and feed it into the model. 
The model is prompted to generate a comprehensive description by integrating speech, background music, acoustic events, and environmental context. 
The generated caption is taken as the final prediction.

For TAG, we process each chunk independently using a binary detection prompt. 
Given a query describing a target event, the model is instructed to determine whether the event exists within the current 60-second chunk. 
The prompt is defined as follows:

\begin{lstlisting}
You are an expert in temporal audio grounding.
1. First determine whether the described event appears in the current audio segment.
2. If yes, output the temporal interval in the format:
   "yes [MM:SS - MM:SS]".
3. If not, output "no".
\end{lstlisting}

We traverse all chunks sequentially and take the first occurrence of a "yes" prediction. 
The corresponding relative timestamp is converted into the global timeline and used as the final prediction.

\section{Case Study}
\label{sec:case study}

\begin{tcolorbox}[
colback=white,
colframe=black,
title=Reduction of Temporal Hallucination and Timestamp Drift,
breakable,
fonttitle=\bfseries,
sharp corners,
left=2mm,
right=2mm,
width=\textwidth
]

\begin{minipage}[t]{0.31\textwidth}
\vspace{0pt}
\textbf{Ground Truth}

\textbf{Audio Range:} Last 4 minutes

\begin{itemize}[leftmargin=*, itemsep=2pt, topsep=2pt]

    \item \textbf{[19:12 -- 20:26]} In the final part of the dialogue, Professor Luo summarizes that legal training has taught him to embrace pluralism like a ``fox,'' rather than insisting on monism like a ``hedgehog.'' He explains that the world is complex, and many conflicts are not between good and evil, but between different forms of ``good,'' which requires balancing competing values and has made him more tolerant.
    
    \item \textbf{[20:27 -- 22:45]} The interviewer mentions that online fans call him the ``Light of Rule of Law'' and asks for his view. Professor Luo modestly responds that this is more of a reminder and aspiration for himself. He further elaborates on the importance of the rule of law and calls on everyone who believes in it to follow rules, so that individual lights may gather into a larger light. The two speakers eventually agree that even ordinary people can become valuable ``small lights'' that influence those around them.
    
    \item \textbf{[22:46 -- 23:27]} The conversation shifts into a promotional closing. The host formally recommends Professor Luo’s new book \textit{Lectures on Criminal Law} and announces that it will be globally launched at a specified time in the ``Jiaoge Pengyou'' livestream. Professor Luo gives his final thanks. After the interview ends, the scene becomes quiet, leaving only low room ambience, with an occasional crisp clicking sound and the sound of an object being gently placed on the table.
\end{itemize}
\end{minipage}
\hfill
\vrule
\hfill
\begin{minipage}[t]{0.31\textwidth}
\vspace{0pt}
\textbf{Gemini-2.5-Pro}

\textbf{Prediction}

\begin{itemize}[leftmargin=*, itemsep=2pt, topsep=2pt]
    \item \textbf{[24:11 -- 26:04]} In the final discussion, Professor Luo reflects on how legal study changed his way of thinking. He contrasts the ``fox'' and the ``hedgehog,'' arguing that real-world conflicts are often not simply between right and wrong, but between different values that both contain some degree of good.
    
    \item \textbf{[26:05 -- 27:01]} The interviewer brings up the nickname ``Light of Rule of Law,'' and Professor Luo responds modestly, saying that it is more like a reminder to himself. He explains that the rule of law depends on people respecting rules, and suggests that even ordinary individuals can contribute a small light.
    
    \item \textbf{[27:02 -- 27:48]} The program moves into a closing promotion for Professor Luo's new book, followed by brief thanks.
\end{itemize}
\end{minipage}
\hfill
\vrule
\hfill
\begin{minipage}[t]{0.31\textwidth}
\vspace{0pt}
\textbf{LAT-Audio (Ours)}

\textbf{Prediction}

\begin{itemize}[leftmargin=*, itemsep=2pt, topsep=2pt]
    \item \textbf{[18:50 -- 20:26]} For ordinary crimes, the statute of limitations is determined with practical considerations such as judicial resources and costs in mind. Professor Luo explains that years of legal training have taught him to embrace pluralism like a ``fox,'' recognizing that the world is shaped by complex and intertwined interests. Many conflicts, he argues, are not between good and evil, but between two different forms of ``good,'' which requires careful balancing.
    
    \item \textbf{[20:26 -- 22:21]} When asked how he feels about being called the ``Light of Rule of Law,'' Professor Luo responds modestly that he dares not fully accept the title, but appreciates the phrase and treats it as a personal reminder and goal. He hopes that he, his peers, and all those who care about the law can each become a light for the rule of law. He further explains that the rule of law requires both just laws and universal compliance. However, many people still treat law merely as a tool and believe that rules are for others. He expresses the hope that everyone who believes in the rule of law, especially those in power, will also submit to rules, so that these lights may converge and illuminate a broader space.
    
    \item \textbf{[22:22 -- 23:27]} The interviewer concludes that even if ordinary people cannot become a great light source, it is still meaningful to be like a candle that influences nearby people. He then formally recommends Professor Luo’s new book \textit{Lectures on Criminal Law} and announces its global launch in the ``Jiaoge Pengyou'' livestream at 8:00 p.m. on July 31.
\end{itemize}

\end{minipage}

\end{tcolorbox}

We conduct a case study on a 23:27-minute audio clip, comparing dense audio captions generated by LAT-Audio, Gemini-2.5-Pro, and the ground truth, with a focus on the final 4 minutes.
We observe that Gemini-2.5-Pro produces timestamps that drift beyond the valid audio duration, with the predicted ending reaching 27:48. Notably, this exceeds the actual audio length (23:27), providing direct evidence of temporal hallucination. Moreover, the entire predicted timeline is globally shifted, resulting in systematic misalignment with the true temporal structure. Although the generated captions roughly follow the high-level semantics, the misaligned timestamps lead to fragmented and partially inconsistent descriptions, particularly in the final segments where the narrative structure becomes incomplete.
In contrast, LAT-Audio maintains strict consistency with the global audio duration and produces temporally coherent segments aligned with the underlying narrative structure. The predicted segments closely align with the true temporal boundaries, with only minor deviations at segment boundaries. Compared to Gemini-2.5-Pro, which exhibits global temporal drift and duration hallucination, LAT-Audio achieves significantly improved temporal alignment and structural consistency while preserving coherent semantic descriptions.
This example highlights how temporal errors accumulate over long durations in baseline models, while LAT-Audio effectively constrains such errors through global-to-local reasoning. These results demonstrate that explicitly modeling global temporal structure and performing iterative evidence-grounded reasoning are crucial for mitigating temporal hallucination and maintaining alignment in long-form audio understanding.

\twocolumn



\bibliographystyle{ACM-Reference-Format}
\bibliography{sample-base}










\end{document}